\documentclass{sig-alternate-2013}

\newfont{\mycrnotice}{ptmr8t at 7pt}
\newfont{\myconfname}{ptmri8t at 7pt}
\let\crnotice\mycrnotice%
\let\confname\myconfname%

\permission{Permission to make digital or hard copies of all or part of this work for personal or classroom use is granted without fee provided that copies are not made or distributed for profit or commercial advantage and that copies bear this notice and the full citation on the first page. Copyrights for components of this work owned by others than the author(s) must be honored. Abstracting with credit is permitted. To copy otherwise, or republish, to post on servers or to redistribute to lists, requires prior specific permission and/or a fee. Request permissions from permissions@acm.org.}
\conferenceinfo{WebSci'14,}{June 23--26, 2014, Bloomington, IN, USA. \\
{\mycrnotice{Copyright is held by the owner/author(s). Publication rights licensed to ACM.}}}
\copyrightetc{ACM \the\acmcopyr}
\crdata{978-1-4503-2622-3/14/06\ ...\$15.00.\\
http://dx.doi.org/10.1145/2615569.2615672}

\usepackage{times}
\usepackage{helvet}
\usepackage{courier}
\usepackage{url}
\usepackage{amssymb}
\usepackage{amsmath}
\usepackage{mathptmx}
\usepackage{graphicx}
\usepackage{hyperref}
\usepackage{subfigure}
\usepackage{multirow}
\usepackage{tabularx}
\usepackage{balance}

\begin{document}
\setlength{\pdfpagewidth}{8.5in}
\setlength{\pdfpageheight}{11in}
\title{Reading the Source Code of Social Ties}

\numberofauthors{3}
\author{
\alignauthor Luca Maria Aiello\\
       \affaddr{Yahoo Labs}\\
       \affaddr{Barcelona, Spain}\\
\email{\large alucca@yahoo-inc.com}   
\alignauthor Rossano Schifanella\\
       \affaddr{University of Torino}\\
       \affaddr{Torino, Italy}\\
\email{{\large schifane@di.unito.it}}
\alignauthor Bogdan State\titlenote{All authors contributed equally to this work, that was done when R. Schifanella and B. State were visiting Yahoo Labs Barcelona.}\\
       \affaddr{Stanford University}\\
       \affaddr{Palo Alto, CA, USA}\\
			\email{{\large bstate@stanford.edu}}
}

\maketitle

\begin{abstract}
Though online social network research has exploded during the past years, not much thought has been given to the exploration of the nature of social links. Online interactions have been interpreted as indicative of one social process or another (e.g., status exchange or trust), often with little systematic justification regarding the relation between observed data and theoretical concept. Our research aims to breach this gap in computational social science by proposing an unsupervised, parameter-free method to discover, with high accuracy, the fundamental domains of interaction occurring in social networks. By applying this method on two online datasets different by scope and type of interaction (aNobii and Flickr) we observe the spontaneous emergence of three domains of interaction representing the exchange of status, knowledge and social support. By finding significant relations between the domains of interaction and classic social network analysis issues (e.g., tie strength, dyadic interaction over time) we show how the network of interactions induced by the extracted domains can be used as a starting point for more nuanced analysis of online social data that may one day incorporate the normative grammar of social interaction. Our methods finds applications in online social media services ranging from recommendation to visual link summarization.
\end{abstract}


\section{Introduction}\label{sec:intro}

The explosion of data from online social media has encouraged the often uncritical adoption of the notion of \textit{social tie} as the atomic interaction quantum of any social network structure. Social ties are usually treated as \textit{a priori} entities, immediately available to the researcher from the graph of online-mediated interactions such as emailing, following on Twitter, or friending on Facebook. The social tie is indeed a powerful abstraction that has allowed researchers to build rigorous models to describe the evolution of social networks and the dynamics of information exchange~\cite{gomez13structure}.

Even though previous research on social networks has explored the \textit{intensity} of social links, as well as their \textit{polarity}~\cite{gilbert09predicting,Leskovecetal:2010}, there is still much to investigate about the \textit{nature} of the social interactions implied by social ties. One way to overcome this limitation is to look into the \textit{content} of social links, the \textit{messages} exchanged between actors.

For these reasons, online \textit{conversations} -- the object of our study -- have emerged as an important domain of research for social link characterization~\cite{dechoudhury09conversations,backstrom2013}. Although the tools to mine the \textit{syntactics} and \textit{semantics} of online conversations are available and have been used extensively~\cite{ramage10characterizing}, to date there is no way to automatically capture the \textit{pragmatics} of communication. From the angle of pragmatics, messages are not just defined by their intensity, structure or topic but can be instead interpreted as \textit{communicative acts} that contribute to the incremental definition of the nature of the social relationship between pairs of individuals. We understand this process of construction of social ties through the lens of Social Exchange Theory~\cite{Blau:1964}, conceiving every dyad as a repeated set of exchanges of different types of non-material \textit{resources} transacted in an interpersonal situation, such as knowledge, social support or manifestation of approval~\cite{Foa:1980}. Being able to describe a conversation in terms of these resources would overcome the limitations of the current representations of social links.

This work gives a contribution in this direction by defining a method to \textit{discover} the types of resources exchanged in a social network and to cluster messages by the type of resource they convey, rather than by their topical aspect. Our algorithm is unsupervised and parameter-free, as the number of clusters is detected automatically and it can be applied to different languages. The algorithm is based on the intuition that in a dyad, social interactions conveying a resource tend to be reciprocated with the same resource type. As an illustration, if two individuals exchange knowledge now, their next exchange will be most likely to also involve knowledge, rather than affection.
This intuition has been validated for a wide range of social interactions in both the offline~\cite{gould02rigins,antonucci90social} and online world. In this work we make the following main contributions:
\begin{itemize}
	\item We propose a novel method to cluster messages based on the type of resource they convey ($\S$\ref{sec:methodology}). Using the bibliophile community aNobii and the photo sharing service Flickr as case study ($\S$\ref{sec:dataset}), our algorithm yields edifying results in detecting meaningful and coherent domains of interaction when compared to a ground truth generated by human coders ($\S$\ref{sec:evaluation}).
	\item We apply our methodology to two datasets of different nature and we observe the spontaneous emergence of three main domains that are identified by as many social exchange processes, namely \textit{status exchange}, \textit{social support}, and \textit{knowledge exchange} ($\S$\ref{sec:interpretation}). 
	\item We provide a framework that enables a direct validation of social theories about well-known interaction types (e.g., status giving) that are difficult to test in practice with a conventional tie representation.
	We take on the issues of tie strength, dyadic interaction over time, and inequality of resource exchange in relation to the different domains of interaction, finding striking regularities across the two datasets ($\S$\ref{sec:analysis}).
\end{itemize}

\section{Related work}\label{sec:related}

\textbf{Online conversations.} A branch of the research studying online conversations has focused on the characterization of the users based on the \textit{conventions} they use, especially in Twitter~\cite{honey09beyond,boyd10tweet}. Correa \emph{et al.}~\citeyear{correa10who} conducted interviews to investigate the correlation between psychological indicators, such as emotional stability and openness to new experiences, with propensity to engage online conversations. On a similar note, Celli and Rossi~\citeyear{celli12role} studied Twitter conversational data, estimated the user emotional stability from the text and correlated it with the tendency to engage conversations.

Conversations around items have been studied also in relation with the \textit{engagement} of users in online communities. De Choundhry \emph{et al.}~\citeyear{dechoudhury09conversations} studied discussions around YouTube videos and estimate the thread interestingness using a random walk model. Backstrom \emph{et al.}~\citeyear{backstrom2013} used a machine learning model to predict the number of entries and the probability for a user to submit a new post in Twitter discussion threads. Harper \emph{et al.}~\citeyear{harper07talk} interpret participation in conversations as a proxy for engagement and, to limit user churn, they proposed to send personalized and familiar invitations to join threads. Budak and Agrawal~\citeyear{budak13participation} studied factors that affect continued user participation in Twitter chats and identify through surveys the distinct dimensions of informational and emotional exchange in messages. Similarly, the application of our method to online datasets finds the emergence of a social support dimension.

A line of work more similar to ours was devoted to the investigation of the properties of conversations. Kumar \emph{et al.}~\citeyear{kumar10dynamics} built a model able to reproduce the size and depth of multi-user conversations in Twitter and Yahoo Groups. Java \emph{et al.}~\cite{java07why} studied the \textit{intent} behind Twitter conversations and based on that they identify different behaviours and types of users. Although informed by hub-authority computation and inspection of communities on the mention graph, their classification is ultimately performed manually. The aspect of emotions conveyed in conversations has been recently studied by Kim \emph{et al.}~\citeyear{kim12feel}. They extracted LDA topics in Twitter conversations, used a framework based on the Plutchik's wheel to assign emotions to them, and analyzed the transitions between emotions in conversations. Similarly to what we find for resource exchange, they verified that a conversation that conveys a certain sentiment tends do it consistently in the following exchanges (``nice words for nice words''). Very recently, more nuanced studies have been done around online conversations, touching upon the concepts of social cohesion and social identity and their implications on group discussion divergence~\cite{purohit14understanding}, and discussing the social power dynamics they contribute to create~\cite{tchokni14emoticons}.

\textbf{Link characterization.} Research on characterization of social links has focused primarily on the concept of \textit{tie strength} on \textit{interaction} networks~\cite{viswanath09evolution,wilson09user}. Gilbert and Karahalios~\citeyear{gilbert09predicting} used a supervised method to predict the tie strength in Facebook using textual, profile and graph structural features. We use apply their framework to characterize the average tie intensity for different social exchange processes ($\S$\ref{sec:analysis:strength}). Xiang \emph{et al.}~\citeyear{xiang10modeling} addressed the same problem with an unsupervised model instead, using a latent variable model based on some profile features, assuming that the higher the profile similarity the higher the strength of the link. Grabowicz \emph{et al.}~\citeyear{grabowicz12social} studied the strength of ties in relation with the communities in the Twitter interaction graphs and identified weak ties as the ones towards community intermediaries. Besides link strength, research has been done on the sign of edges in network with positive and negative links (e.g., Slashdot) mostly in the direction of link sign prediction~\cite{Leskovecetal:2010}. So far, little attention has been devoted to characterize the type of social links according to sociological dimensions, and recent work on the accommodation of linguistic styles according to power differentials provides an example of the intellectual opportunities now available at the intersection of social theory and conversational data~\cite{danescu12echoes}. A step to fill that gap has been recently done by Bramsen \emph{et al.}~\citeyear{bramsen11extracting}, who have introduced a supervised approach to identify social power relationships in social dyads using ad-hoc texual features. Our method provides a means for the discovery of multiple kinds of social exchange in an unsupervised way, rather than limiting the interaction to status exchange.

\section{Methodology} \label{sec:methodology}

\subsection{Problem Definition}

The general problem we address is defined as follows.

\noindent\textbf{Input}: a population of users $U$ and a set of messages $M$ where each message $m_{u,v}^t \in M$ is a textual communication between source $u \in U$ and destination $v \in U$ at time $t$.

\noindent\textbf{Output}: a probabilistic clustering of messages in $M$ with probability of a message $m$ to be assigned to cluster $D$ being $p(m,D) \geq 0$.

The novel aspect of the method is the nature of the clusters in output, that do not group together messages based on their topical aspects, but instead according to the type of social exchange those messages convey. The algorithm is composed by four phases: 1) preprocessing and distillation of the raw text messages, 2) clustering of messages in buckets according to their textual similarity, 3) creation of a conversation graph that models the transitions between buckets during social interactions, and 4) extraction of dense portions of the conversation graph through a community detection algorithm. We will describe in details each step in the following sections. 

\subsection{Preprocessing}

We apply to the raw text a series of filters commonly used in information retrieval. The filters include the removal of non- alphanumeric strings, stopwords, and very frequent and infrequent terms, namely those who appear in more than $60\%$ and less than $1\%$ of the corpus. To reduce inflected forms to their root we apply a stemming algorithm. After a tokenization phase, a message representation is expanded with the insertion of bi-grams and tri-grams to take into account the discriminative power of sequences over single terms (e.g., the bigram \emph{``great shot''} is more informative than the individual terms \emph{great} and \emph{shot}).

The adoption of $n$-grams can lead to an explosion of the dimensionality of the feature space and, in a practical scenario, an upper bound based on term frequency is needed. We consider only the most frequent 10,000 $n$-grams with $n \in [1...3]$ and we filter out messages that do not contain elements in that vocabulary (less of $1.5\%$ for both corpora).

The vector of stemmed terms representing the messages are stacked in a \emph{term-document matrix} $\Gamma_{m \times n}: w_{ij}$ where $m$ is the number of terms in the vocabulary and $n$ is the number of messages in the corpus. A generic element $w_{ij}$ reflects the importance of the corresponding term $i$ with respect to the semantics of message $j$ and it is calculated with a standard TF-IDF weighting scheme with sublinear TF scaling. This matrix is the only input to the next stages of the pipeline.

\subsection{Message Bucketing}\label{sec:bucketing}

Modern social media convey a huge volume of information through the interactions between users. Modeling these dynamics as message-to-message communication process can raise practical issues due to the dimensionality of the data flow. Moreover, conversations are often characterized by variations of recurrent patterns that use similar sentences and words for conveying the object of the conversation. For instance, greetings in an online community could be coded in different variations (e.g., \emph{``Hi, how are you?''} or \emph{``Hello, how do you do?''}). These observations suggest the possibility to model conversations not as transitions between single messages but instead as transitions between classes of homogeneous messages. 

To this extent, we leverage a probabilistic generative model based on a low rank \emph{Non-negative Matrix Factorization (NMF)} method to cluster messages in coherent groups according to their textual content. We name these homogeneous clusters \emph{message buckets}. NMF has been successfully used in document clustering~{\cite{Xu2003} and topic detection tasks~\cite{DBLP:journals/corr/abs-1204-1956} and it allows a part-based representation where a document is modeled as an additive combinations of topics vectors due to the non-negativity constraint. In a text mining framework, this property differentiates NMF from other existing matrix decomposition approaches like \emph{Singular Value Decomposition (SVD)} or \emph{Principal Component Analysis (PCA)} that force a document to belong to a single topic or are able to recover only the span of the topic vectors instead of the topic vectors themselves~\cite{DBLP:journals/corr/abs-1204-1956}.

The NMF model is able to factor the previously defined non-negative \emph{term-document matrix} $\Gamma_{m \times n}: w_{ij}$ into two matrices $W_{m \times k}$ and $H_{k \times n}$ such that $\Gamma = W H + e$, where $e$ is a $m \times n$ matrix of approximation errors, and where $k \ll m$.
In short, entries of the matrix $W$ represent the probability of each of the $m$ terms to belong to each of the $k$ buckets, whereas the matrix $H$ embeds the probability of each bucket to include each of the $n$ messages. This approach fits well into the assumption that a message can convey multiple informational units and then belong to different buckets. The matrix decomposition enables the definition of two functions:

\begin{enumerate}
	\item $\mathcal{B}(H, m_i)$, maps a message $m_i$ into the set of most representative buckets,
	\item $\mathcal{T}(W, b_i, n)$, maps a bucket $b_i$ into the set of $n$ most characterizing terms.
\end{enumerate}

The choice of the number of buckets $k$ is generally application-dependent. Many different methods for evaluating the optimal number of underlying components $k$ have been developed in this context~\cite{DBLP:journals/bioinformatics/FogelYHL07}. In particular, we use an iterative approach that selects the $k$ that minimizes the \emph{Frobenius norm} of the error matrix $e$. 

\subsection{Building The Conversation Graph}

To shape the conversational aspect of between-user interactions, we introduce the concept of \emph{Conversation Graph} -- a weighted directed graph where nodes are buckets and edges represent transitions between buckets determined by the conversational flow. Intuitively, an edge $(i,j)$ captures the following notion: given a message from ego to alter classified in bucket $i$, what is the likelihood that alter will reply back to ego with a message in bucket $j$? 

Consider a dyad involving users $u$ and $v$ and the time-ordered sequence of messages between them, that is part of the algorithm's input. We define a \emph{transition} $ t_{uv} = (m_{uv}^{t_0}, m_{vu}^{t_1}), t_0 < t_1$ to be a pair of two consecutive mutual messages sent between user $u$ and $v$. Similarly to web browsing session analysis, a threshold on the the elapsed time between messages could be used to avoid considering transitions between messages sent with a big temporal gap between each other, and therefore likely to be part of two separate conversations. However, such threshold could vary significantly depending on the medium (e.g., longer time could elapse in email conversations than in instant messaging) and even on the specific user pair, so to keep our approach as general as possible we do not introduce this filtering step.

With this definition in mind, we create the Conversation Graph following these steps:
\begin{itemize}
\item For each pair of users $u$ and $v$ we extract the set of transitions $T_{uv}$ between them.
\item For each transition $t \in T_{uv}$, with $t = (m_x, m_y)$, we derive the sets of most representative buckets using the function defined in $\S$\ref{sec:bucketing}. We obtain $\mathcal{B}(H, m_x) = B_i$ and $\mathcal{B}(H, m_y) = B_j$.
\item $\forall b_i \in B_i$ and $\forall b_j \in B_j$ with $b_i \neq b_j$ we add a directed edge $b_i \rightarrow b_j$ with weight $w_{ij} \in [0,1]$ that is proportional to the probability of the messages $m_x$ and $m_y$ to belong to the corresponding buckets. Such weights are extracted from the matrix $H$ computed in $\S$\ref{sec:bucketing}. 
\end{itemize}

The process of construction of the Conversation Graph is illustrated in Figure~\ref{fig:transition_example}. In the example, a user $u$ writes a message $m_1$, belonging to bucket $A$, to user $v$ and gets as reply a message $m_3$, belonging to bucket $B$. This interaction implies that there is a conversational transition from messages in $A$ to messages in $B$, and a directed arc between them is created accordingly.

\begin{figure}
\center
 \includegraphics[width=0.65\columnwidth]{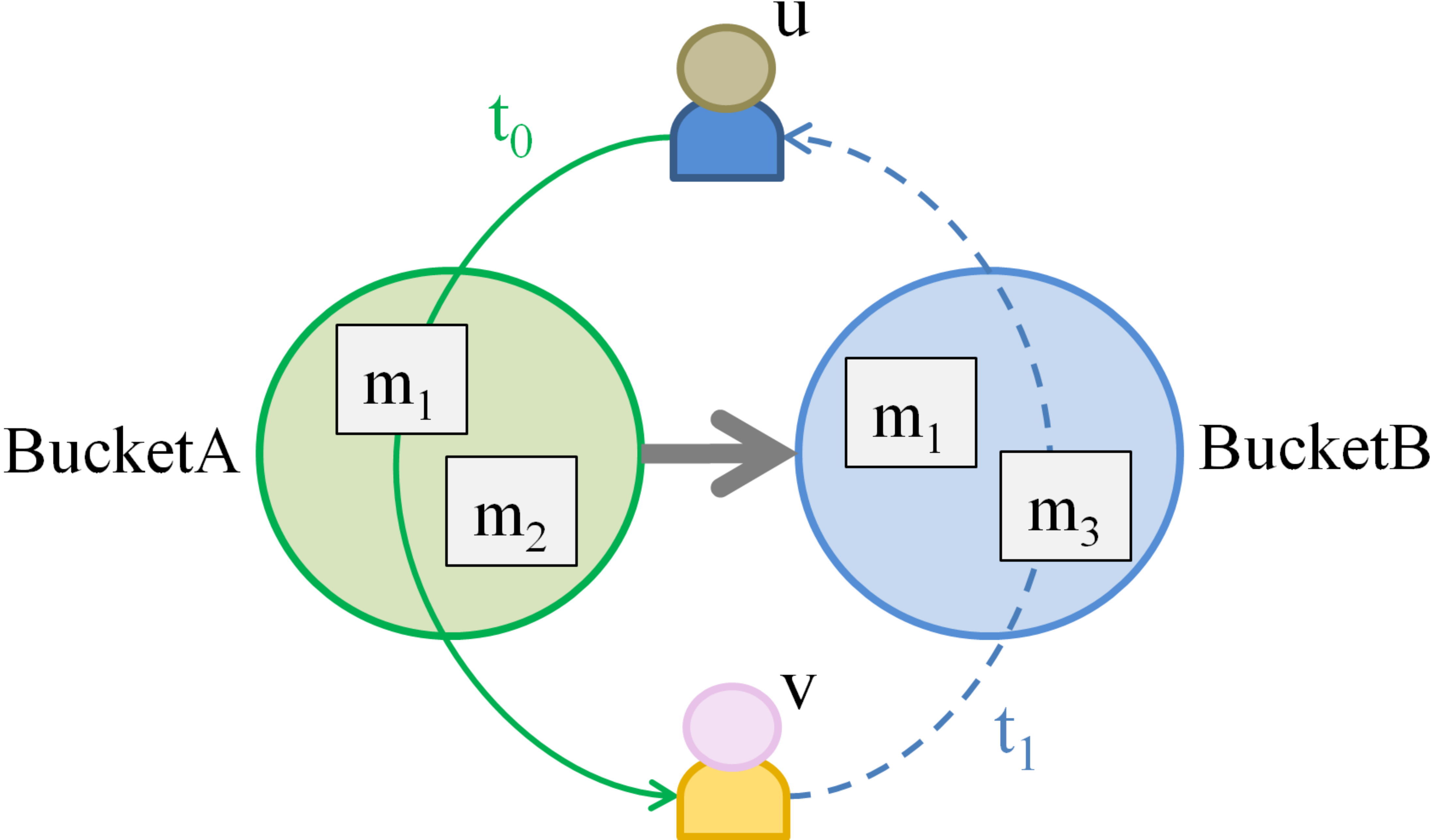}
 \footnotesize 
 \caption{Example of construction of the Conversation Graph from conversation data.}
 \label{fig:transition_example}
\end{figure}

\subsection{Extracting Domains of Interaction}\label{sec:community}

The Conversation Graph shapes the transition between classes of coherent messages during social interactions. We conceive these interactions as the realizations of underlying processes of social resources exchange and we assume that a message that conveys a certain type of resource will most likely get a reply that conveys the same resource type.

In offline social networks the propensity to reciprocal interactions has been derived as a theoretical necessity in the exchange of status~\cite{gould02rigins} and has been shown to exist empirically in the case of social support~\cite{antonucci90social}. Moreover, in the online world reciprocity has been found to exist for a wide range of social interactions~\cite{kim12feel}. 

Our work does not make the assumption that reciprocity is ubiquitous in human interactions. Rather, we follow previous work in assuming that if reciprocation is observed, then the reciprocal interaction will be likely in the same social domain (e.g., of status exchange, or of social support, etc.) as the initial interaction. 
For instance, we would expect a person who receives social support for the loss of a grieving relative (``I'm sorry for your loss'') to reply in kind (if at all) with another social support interaction (``Thank you for being a good friend'') rather than a status-exchange interaction (``You're such a great photographer!''). Indeed, we verify this assumption in our experimental setting, which yields coherent domains of interaction for two independent datasets (see $\S$\ref{sec:evaluation}).

Under this interpretation, highly-clustered parts of the Conversation Graph aggregate buckets that carry homogeneous patterns of social exchange and will have fewer edges connecting them to the rest of the graph. This scenario is consistent with the most common definition of graph \textit{community}~\cite{fortunato10community}, therefore network community detection algorithms could be applied to the Conversation Graph to discover these dense areas. In our experiments we use the the Spinglass algorithm~\cite{ReichardtBornholdt:2006} available in the igraph library.

We name \textit{Domains of Interaction} (DoIs) the communities given as output by the community detection algorithm, as in our conception they contain messages that belong to a domain in which the resources exchanged during interactions tend to be homogeneous. The final output of the community detection step is a fuzzy assignment of messages to a set of DoIs: every message is assigned to every DoI that includes at least one bucket containing that message, with a probability equal to the maximum probability of the message belonging to one of those buckets.

The algorithm is fully unsupervised, but it does not allow us to assign labels to the extracted domains. The interpretation of the nature of the domain is admittedly a task that is hard to accomplish automatically and social-scientific input is necessary to provide \textit{qualitative} insights into the algorithm's findings, combining the emerging clustering with social theory. Next ($\S$\ref{sec:dataset}) we present the details of the two datasets we used to test our method and after that ($\S$\ref{sec:interpretation}) we describe the application of our method to them and the process of interpretation of the domains we obtained.

\section{Datasets}\label{sec:dataset}

We test our framework on datasets extracted from two social media: aNobii, a website for book lovers, and Flickr, the popular image sharing website. Both have similar mechanisms for the creation of social connections: social ties are directed and, similarly to the ``following'' relation available in other mainstream social media, they allow users to receive all the updates of the profiles they are linked with. Social links can be created towards any other user, without the need of any authorization. Peculiar aspects of the two networks are discussed next.

\noindent\textbf{aNobii.} User profiles in aNobii are centered around a personal digital library containing the titles the users have read. The main channel of interaction is the public messaging activity: every profile page contains a public \textit{shoutbox} where any user can leave a message and see the messages written by others. It is common practice for pairs of users to engage conversations by writing on each other's shoutbox. 
We use a public aNobii dataset recently released to the public~\cite{aiello12people} and we model conversations through a \textit{communication graph} where nodes are users and directed arcs represent the messages exchanged between them.
Users write in different languages, but the biggest community is the Italian one, accounting for around $35\%$ of the user base and for $76\%$ of the message traffic. A cross-language analysis is outside the scope of this work, so we focus on the Italian community only. We consider all the messages (around $1M$) exchanged over the $\sim545k$ unique pairs of Italian users between year 2006 and end of year 2011.

\textbf{Flickr.}
Differently from aNobii, Flickr does not provide any tool for sending direct public messages between users, therefore communication is mainly mediated by the activity of photo commenting: a pair of users can either initiate a communication thread by commenting under a user's photo or by writing comments on each other's photos. Although both are possible, we consider the second option only. This choice appears reasonable first because, even if the direct target of the comment is the photo, its main recipient is always the photo owner, who is the only one being explicitly notified of the new comment. Additionally, since a Flickr persona is defined mainly by its photos, writing on a photo is a quite common practice to convey a message directly to the owner, as observed in previous studies on item-mediated communication~\cite{mitrovic12dynamics}. On the other hand, the first option is not practical because identifying the communication flow (who is writing to whom) in a thread with potentially many commenters is an arduous task~\cite{ritter10unsupervised}, and the assumption of a message being always delivered in broadcast to all the thread participants would be an unacceptable oversimplification.

Similarly to aNobii, we model the interactions with a communication graph, arcs of which go from the commenters to the owners of the commented photos. To get a sample of conversations, we randomly selected 100k anonymized user pairs who commented on each other's photos at least once. For each of these pairs we collect the full history of their comment exchange, getting around 2M messages in total.

\begin{table}[t]
\small
\centering
  \begin{tabular}{c|c|c|ccc}
		               & \textbf{Users}  & \textbf{Conversations}  & \textbf{$conv_{len}$} & \textbf{$msg_{len}$} \\
	 \hline
	 aNobii & 62,235 & 545,656        & 1.75 (1)              & 18.75 (13)           \\
	 Flickr & 95,397 & 100,000        & 10.84  (3)            & 6.90 (5)             \\
	\end{tabular}
 \caption{Size of the two datasets, average value and median (in paranthesis) of message length (number of tokens) and conversation length (number of messages).}
\label{tab:dataset}
\end{table}

A summary of some basic quantities of the two dataset is provided in Table~\ref{tab:dataset}. Although they share commonalities, Flickr and aNobii are quite different domains for scope and norms of interaction, beginning with the different ways of exchanging messages (direct vs. item-mediated). This difference is already surfaced by basic statistics such as the average message length, that is way lower in Flickr. Given the short length of Flickr messages, we assume the likelihood of a message conveying multiple resources will be low. For this reason, in the case of Flickr we don't consider a probabilistic assignment of message to buckets, but instead we assign each message to the most likely bucket.

\section{Extraction of Domains of Interaction}\label{sec:interpretation}

\begin{figure*}[t]
\centering
\subfigure[aNobii]{\includegraphics[width=0.75\columnwidth]{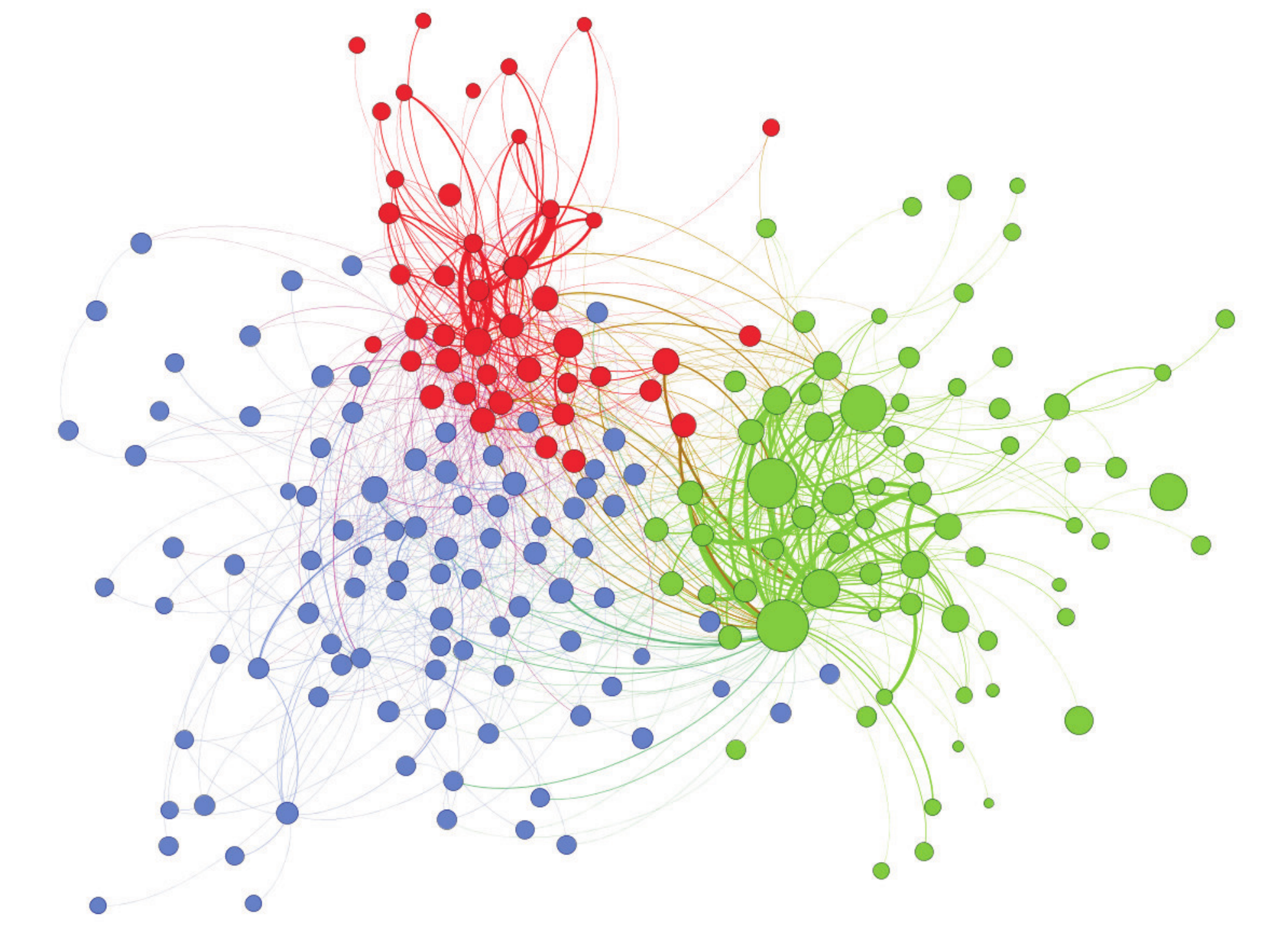}}
\hspace{5mm}
\subfigure[Flickr]{\includegraphics[width=0.70\columnwidth]{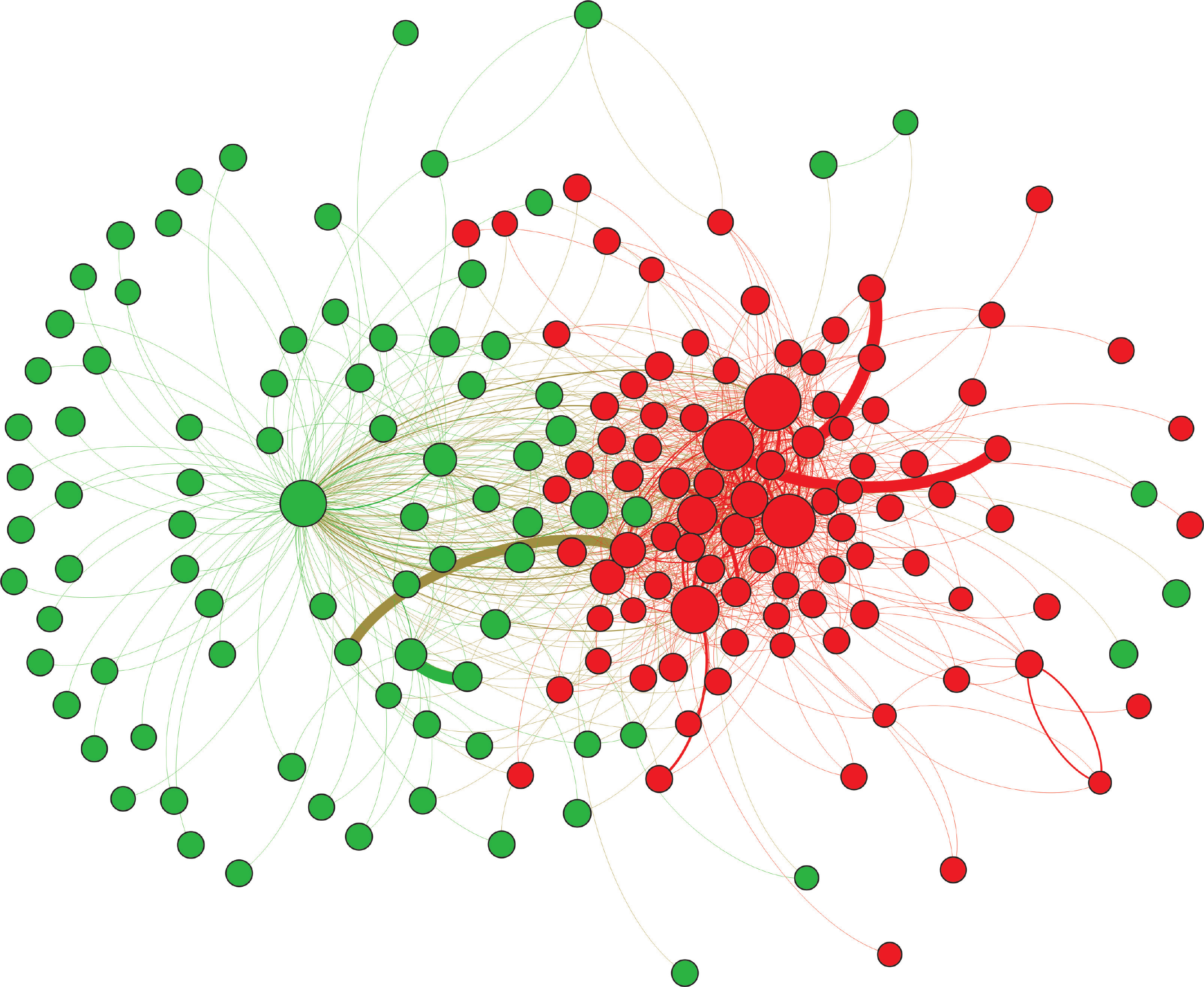}}
\caption{Conversation Graphs. Nodes represent message buckets and edges the transitions between them. Colors encode the Domains of Interaction of \textit{Social Support} (red), \textit{Status Exchange} (green), and \textit{Knowledge Exchange} (blue). Edge thickness is proportional to the weighted number of transitions between buckets and the size of nodes is proportional to the number of messages in the bucket. Graph layout is arranged according to the Yifan-Hu algorithm, edges with low weight are not displayed. Graph drawing powered by gephi.org.}
\label{fig:transitiongraph}
\end{figure*}

We apply the methodology described in $\S$\ref{sec:methodology} to both datasets, thus obtaining a mapping of each message to DoIs. The Conversation Graph of message buckets is depicted in Figure~\ref{fig:transitiongraph}. The optimal number of buckets $k$ we found for aNobii and Flickr (as determined by the error computed on the output of the NFM algorithm described in $\S$\ref{sec:bucketing}) are $350$ and $250$ respectively, but we also verified the DoIs boundaries to be resilient to significant changes of $k$. The Spinglass community detection algorithm yields three distinct communities in aNobii and two in Flickr.

For illustration we show in Table~\ref{tab:doiwords} the most representative terms for the five domains, selected by summing the weights of the terms in each bucket. To get a first interpretation of their nature, we have shown the most frequent terms and a sample of messages from each cluster to a sociologist. This inspection suggests that the three domains in aNobii correspond to as many fundamental processes of social exchange: \textit{Knowledge exchange}, \textit{Status exchange}, and \textit{Social Support}. Accordingly, in Flickr analogous domains emerge, with the exception of the one related to knowledge exchange.

Albeit we do not consider this classification to be exhaustive of all the possible types of interaction, and many other kinds of resources could spawn other forms of social exchange~\cite{Foa:1980,DerksBosGrumbkow:2007}, we see social exchange processes occurring in these domains as essential for the development of most social ties, especially in the wide context of online, task-oriented communities.

Next we report an interpretation of the three domains found, along with some representative messages.

\begin{table}[t]
\begin{center}
    \footnotesize
    \begin{tabular}{c|c|c}
		  & \textbf{DoI}                     & \textbf{Most representative tokens} \\ 
		  \hline
		  \multirow{3}{*}{\rotatebox[]{90}{aNobii}} & \textit{Status}  & neighbor $\cdot$ library $\cdot$ like $\cdot$ congratulations $\cdot$ visit \\
		  
		  & \textit{Support}  & wish $\cdot$ dear $\cdot$ good day $\cdot$ greeting $\cdot$ friend $\cdot$ soon\\
		  & \textit{Knowledge}    & book $\cdot$ read $\cdot$ know $\cdot$ think $\cdot$ advice\\
		  \hline
		  \parbox[t]{2mm}{\multirow{2}{*}{\rotatebox[]{90}{Flickr}}} &	\textit{Status}    & beautiful $\cdot$ wonderful $\cdot$ photo $\cdot$ capture $\cdot$ great shot \\
		  & \textit{Support}      & guy $\cdot$ haha  $\cdot$ year $\cdot$ happy $\cdot$ lol $\cdot$ day \\
    \end{tabular}
 \caption{Selection of the most representative words (and $n$-grams) in the DoIs, according to the NMF weighting. For the sake of presentation, stems used in NMF are inflected and then translated from Italian in the case of aNobii.}
 \label{tab:doiwords}
  \end{center}
	
\end{table}

\subsection{Status exchange}
In most social contexts the possession of resources is often non-uniformly distributed across the actors. In a task-oriented online social network, for example, some people can own more items (photos, books, social contacts) than others, and the distribution of item possession is usually heavy-tailed. According to the Power-Dependence Theory~\cite{cook78power}, this heterogeneity of resource endowments in a dyadic relationship leads to \textit{power imbalances} and a situation of power inequality induces a behavior that may bring the relationship closer to a more balanced state. Among the power-balancing mechanisms, \textit{status giving} is a way in which a low-power actor may attempt to lessen their dependence on a more powerful partner~\cite{state11power}.
In practical terms, status giving is often instantiated in messages displaying appreciation, esteem, or admiration sent to social partners with higher power. Expressions of status giving in aNobii and Flickr are often related to the display of admiration for other people's books or photo collections, such as ``very interesting library'' or ``excellent shot''. In both cases, besides the appreciation for the showcased items in the user profile (e.g., ``Beautiful scene and well captured by you''), the explicit declaration of the act of creating a new social link is also a communicative act that implies status giving (e.g., ``Hi, interesting profile, I added you as my neighbor''). This is coherent with the notion of \textit{prestige} in social network analysis being related to the centrality of an actor in the social graph~\cite{wasserman94social}. Symmetrically, acknowledging the attention received (e.g., ``Thank you very much for your visit'') is also a way to express gratitude that is part of the status exchange ritual.

\subsection{Social Support}
Many everyday interactions have comparatively little to do with the previously described process of status giving. Indeed, many interactions seem inconsequential: greetings, chit-chat with a coworker, gossiping with a friend, wishing a person well, or discussing everyday problems with a sibling. These usually-minute exchanges between individuals form the essential structure of social interactions, that of social support, a basic process of friendship through which one partner provides emotional valuation to another.

A first attempt of generalization of the concept comes from House et al.~(\citeyear{Houseetal:1988}), who define social support as ``the positive [...] aspects of relationships, such as instrumental aid, emotional caring or concern, and information.'' This wider notion of support has been studied in web-mediated interaction~\cite{Rheingold:1993} and in the context of urban areas~\cite{wellman90different}, in which companionship and minor emotional aid are part of the daily interpersonal interactions.

In the datasets we consider, expressions of social support are varied, ranging from sending good wishes (``Bye, I wish you a merry Christmas and a happy 2012'') to colloquial chat (``My dear, I found you also here! How are you doing?'', ``sooo soo cute! you looked good as a baby''), jokes and laughter (``lol, thanks! Right back at ya!''). In Flickr especially this seems to reflect quite well the type of interaction happening in social groups (as opposed to the topical ones) that has been detected in previous work~\cite{grabowicz13distinguishing}. 

\subsection{Knowledge exchange}
Often the main resource being exchanged on a social media platform is knowledge related to the platform's orientation: technical knowledge on stackoverflow.com, knowledge about music on last.fm, or book-related knowledge on aNobii. Even though we have no direct way of gauging the nature and quality of the information particular individuals possess, we can observe the act of sharing one's knowledge or opinions with others (e.g., ``To have a general overview on this, my advice is to read the introductory books by Todorov and Baudot'', ``I appreciate the author, he is still young but very good and it reminds me the Baricco's writing style''). Given that knowledge about books represents the main resource available to aNobii users, they are expected to provide ample evidence of what books they know about in their messages sent to others. We therefore identify as displaying (or asking for) knowledge the language where works of literature figure prominently. 

Even if some traces of knowledge exchange can be found in Flickr comments (e.g., exchanging opinions on camera models), these kind of conversations appear to be very rare if compared to the other social interaction types and our algorithm does not detect them as a separate domain. Intuitively, this is due to the focus of the platform: while the core purpose of aNobii is to foster discussions about books, Flickr facilitates more the users to focus on the aspect of multimedia exploration and discovery rather than on discussions driven by specific topics.

\section{Evaluation}\label{sec:evaluation}

A first inspection of the algorithm output made by an expert of the domain allowed us to label each community according to the most likely DoIs informed by the literature. We hypothesize that the clusters found coincide with the domains described in $\S$\ref{sec:interpretation}. To verify that, we resort to human evaluation: we produce labeled corpus of messages as ground truth ($\S$\ref{sec:results:groundtruth}) and we match it with the automatically extracted DoIs to check their quality ($\S$\ref{sec:results:evaluation}).

\subsection{Ground truth extraction}\label{sec:results:groundtruth}

To gauge the quality of the output of our method, we produce an editorial ground truth to assess whether a message is assigned to the proper DoI. Two editors read a sample of 1,000 randomly selected messages from each website and label them according to the DoI they belong to. To help the editors with their decision, we provided a description of the DoIs, similar to what is presented in $\S$\ref{sec:interpretation}, and a set of guidelines to perform the assignment. We summarize the guidelines as follows:
\begin{table}[t]
\begin{center}
    \small
    \begin{tabular}{p{7cm}|p{0.8cm}}
		  \textbf{Message} & \textbf{DoI}\\ 
		  \hline
		  Have a good weekend my dear. & \textit{Sup} \\ 
		  \hline
		  Hi! very interesting library! I added you as my neighbor. & \textit{Sta} \\
		  \hline 
		  No, haven't read it, but I read some good reviews. & \textit{Kno} \\ 
		  \hline
		  Of course I remember you, how are you? You've a very good library!  & \textit{Sup} \textit{Sta} \\ 
		  \hline
		  Merry Christmas to you! Yes, I've really enjoyed the last one from Pennac. & \textit{Sup} \textit{Kno} \\ 
		  \hline
		  It's a pleasure to add you back. I see you like Sci-fi! & \textit{Sta} \textit{Kno} \\ 
		  \hline
		  Hi, hope you're doing well. Your latest reviews are good! I just started Harry Potter and I'm loving it.& \textit{Sup} \textit{Sta} \textit{Kno} \\ 
		  \hline
		  Yes, but today is Monday & ? \\
    \end{tabular}
 \caption{Examples of aNobii messages (tr. from Italian) along with the domain of interaction they belong to, according to the editorial labeling process ($\S$\ref{sec:results:groundtruth}).}
 \label{tab:groundtruthexamples}
  \end{center}
\end{table}
\begin{itemize}
	\item A message belongs to the \textit{status exchange} DoI when: it contains explicit appreciation for the profile or activity of another user (e.g., his reviews, his tastes, the size of the library or the quality of photos); it announces the creation of a social tie; it points out commonalities between users and taste compatibility; it acknowledges the attention received from others.
	\item A message belongs to the \textit{social support} DoI when: its main purpose is to greet or welcome someone to the website; it explicitly expresses affection or attachment; it contains wishes, jokes, or laughter.
	\item A message belongs to the \textit{knowledge exchange} DoI when: its purpose is to share information and personal experience about books, reading, or related events such as book lovers' meetups; it asks for opinions or suggestions; it displays knowledge of the literary field; it asks for recommendations or suggestions. 
\end{itemize}

If the message appears to be a concatenation of two or more messages that could be standalone messages belonging to different DoIs, then they should be marked with multiple labels. The three DoIs we analyze are \textit{not} supposed to cover all the possible communication patterns in the social network, so no label is given when the message does not seem to belong to any of those reported above. We find that the portion of unlabeled message is quite small ($<10\%$), supporting the intuition that the three DoIs under examination include the vast majority of social interaction types in the social network. The inter-label agreement between the two labelers, measured as Fleiss' Kappa, is 0.70, indicating substantial agreement. Examples of aNobii messages with different labels are displayed in Table~\ref{tab:groundtruthexamples}.

\subsection{Validation}\label{sec:results:evaluation}

\begin{figure}[t]
\begin{center}
 \includegraphics[width=\columnwidth]{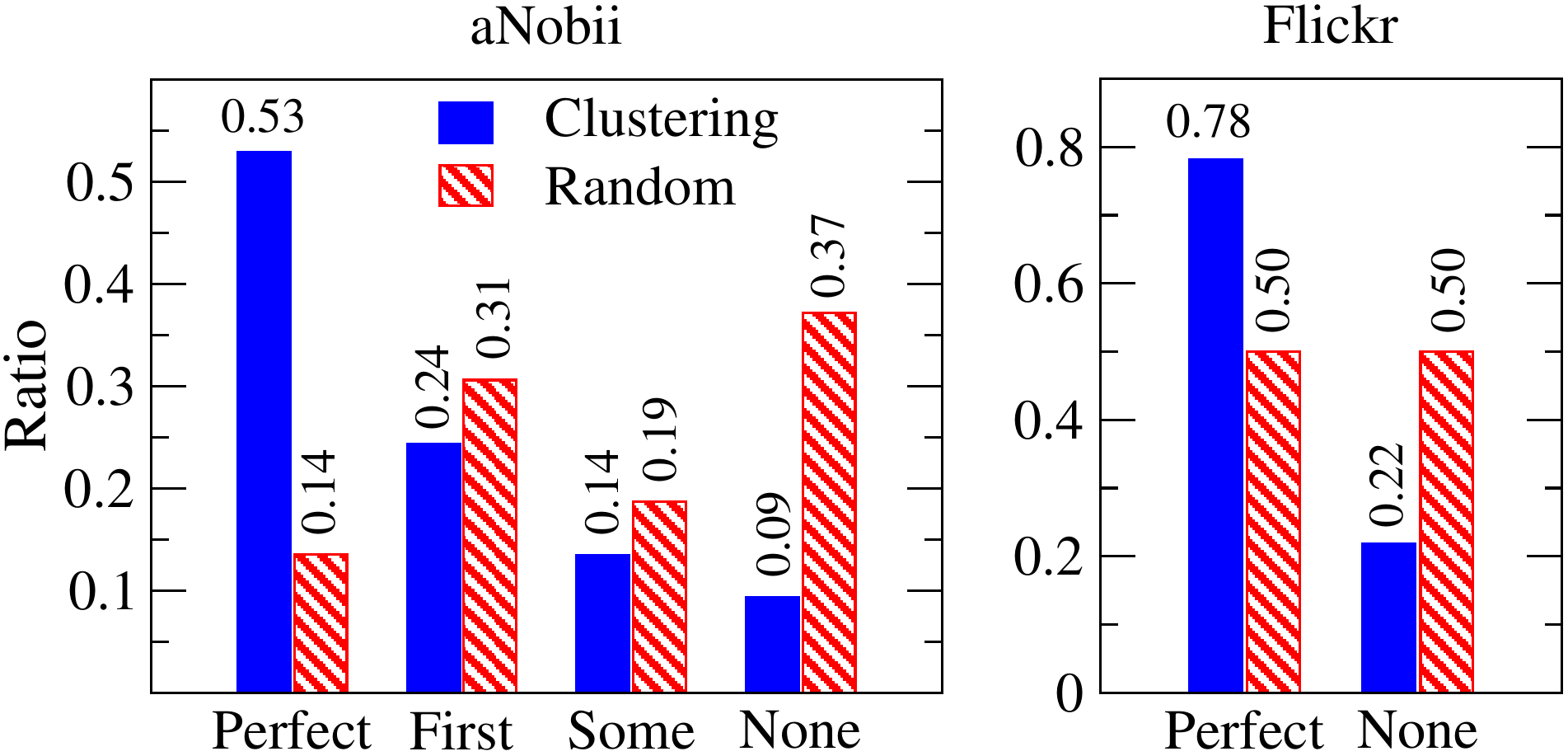}
\end{center}
 \footnotesize 
 \caption{Match of clustering algorithm against the ground truth. Ratio of instances with different level of matching, together with the same values computed for a random clustering.}
 \label{fig:groundtruthmatch}
\end{figure}

The soft clustering method assigns every message to at least one of the detected DoIs. When a multiple assignment occurs, the DoIs are sorted by the probability of membership. For every message in the ground truth corpus, we compare the sorted list of automatically extracted DoIs ($\mathcal{L}^{DoI}_{algo}$) with the set of DoIs assigned by the editors ($\mathcal{S}^{DoI}_{edit}$). We compute the match between them as follows. We consider a \textit{perfect match} when $set(\mathcal{L}^{DoI}_{algo}) \equiv \mathcal{S}^{DoI}_{edit} $; a \textit{first match} occurs instead when $\mathcal{L}^{DoI}_{algo}[0] \in \mathcal{S}^{DoI}_{edit}$; a \textit{partial match} occurs when there is no first match but $set(\mathcal{L}^{DoI}_{algo}) \cap \mathcal{S}^{DoI}_{edit} \neq \varnothing$; there is \textit{no match} when the two sets are disjoint. To compare the obtained results with a baseline, we also compare the ground against random lists of labels ($\mathcal{L}^{DoI}_{rand}$) assigned to every message.

As described in $\S$\ref{sec:dataset}, we run a soft clustering on aNobii and a hard clustering on Flickr, therefore in Flickr only the perfect match and no match categories apply. The results are shown in Figure~\ref{fig:groundtruthmatch}. In aNobii the proportion of perfect matches is close to $53\%$, and the sum of perfect matches and first matches reaches around $77\%$, which is an extremely good results for an unsupervised method dealing with multiple labelled message instances. The second series of bars shows the ratios for the random model, in which the number of perfect matches drops to $14\%$ and the number of wrong matches rises up to almost $40\%$. A similar performance is obtained in the case of Flickr, with an accuracy of $78\%$. In aNobii, the average clustering precision (i.e, ratio between the number of correctly assigned DoIs and the number of automatically detected DoIs, per message) is around $0.76$ for the clustering algorithm, $78\%$ higher than the random case, whose accuracy is around $0.45$.

\section{Analysis}\label{sec:analysis}

The possibility of automated extraction of Domains of Interaction opens opportunities in the field of computational social science, as it allows the social analyst to quantitatively check theories specific to defined sociological categories (e.g., status giving) directly against the detected domain. To illustrate this opportunity we study the structural and evolutionary properties of the communication graphs denoted by each DoI we extracted, namely the subgraphs of the communication networks induced by the edges over which the messages belonging to that specific DoI are delivered.

\begin{table}[t]
\begin{center}
    \footnotesize
    \begin{tabular}{c|c|cccc}
		          \multicolumn{1}{c}{}  & & \textbf{Nodes} & \textbf{Edges} & \textbf{Messages} & \textbf{Reciprocity}\\ 
		  \hline
		  \multirow{3}{*}{\rotatebox[]{90}{aNobii}} & \textit{Status}       & 0.877 & 0.753 & 0.552 & 0.861 \\
		  & \textit{Support}      & 0.726 & 0.400 & 0.409 & 0.783 \\
		  & \textit{Knowledge}    & 0.861 & 0.594 & 0.648 & 0.798 \\
			\hline
		  \multirow{2}{*}{\rotatebox[]{90}{Flickr}}  & \textit{Status}       & 0.821 & 0.660 & 0.368 & 0.757 \\
		  & \textit{Support}      & 0.910 & 0.639 & 0.501 & 0.737 \\
  
    \end{tabular}
 \caption{Statistics about the subgraphs of the communication network induced by the DoIs. Number of nodes, edges, and messages are divided by the same quantities in the full datasets.}
 \label{tab:doinets}
  \end{center}
\end{table}

\subsection{Coverage and reciprocation}
The first question that comes naturally is about how much the different DoIs spread over the communication network. Statistics on the size and link reciprocity of each DoI graph are reported in Table~\ref{tab:doinets}. In the case of aNobii, the difference in the number of edges involved is quite significant, although not very unbalanced in terms of nodes, with the status exchange domain spanning over $75\%$ of the links and social support covering only $40\%$ of them. Consistently with the main purpose of the service, the overall number of messages is instead imbalanced towards the knowledge exchange domain (60\% of ties have a component of domain-related information transmission). In Flickr, instead, the proportion of edges in each domain in more balanced (about $66\%$ for status and $64\%$ for support).
The overall reciprocity in the actors' behavior over the span of a conversation, computed as the ratio of reciprocated messages between two endpoints (disregarding their temporal order) is reported as well. In aNobii, most conversations involve a relatively balanced exchange of messages, on average there being $0.834$ messages sent one way in a conversation for every one message sent in the other direction. The same measure is the highest ($0.861$) for status exchange, likely a reflection of social norms imposing the ritualized reciprocation of status exchange~\cite{gould02rigins}. A similar pattern is found for Flickr. Conversely, both social support and knowledge exchange are less balanced, suggesting slightly more lopsided relationships in these domains of interaction.

\begin{figure}[t]
 \includegraphics[width=\columnwidth]{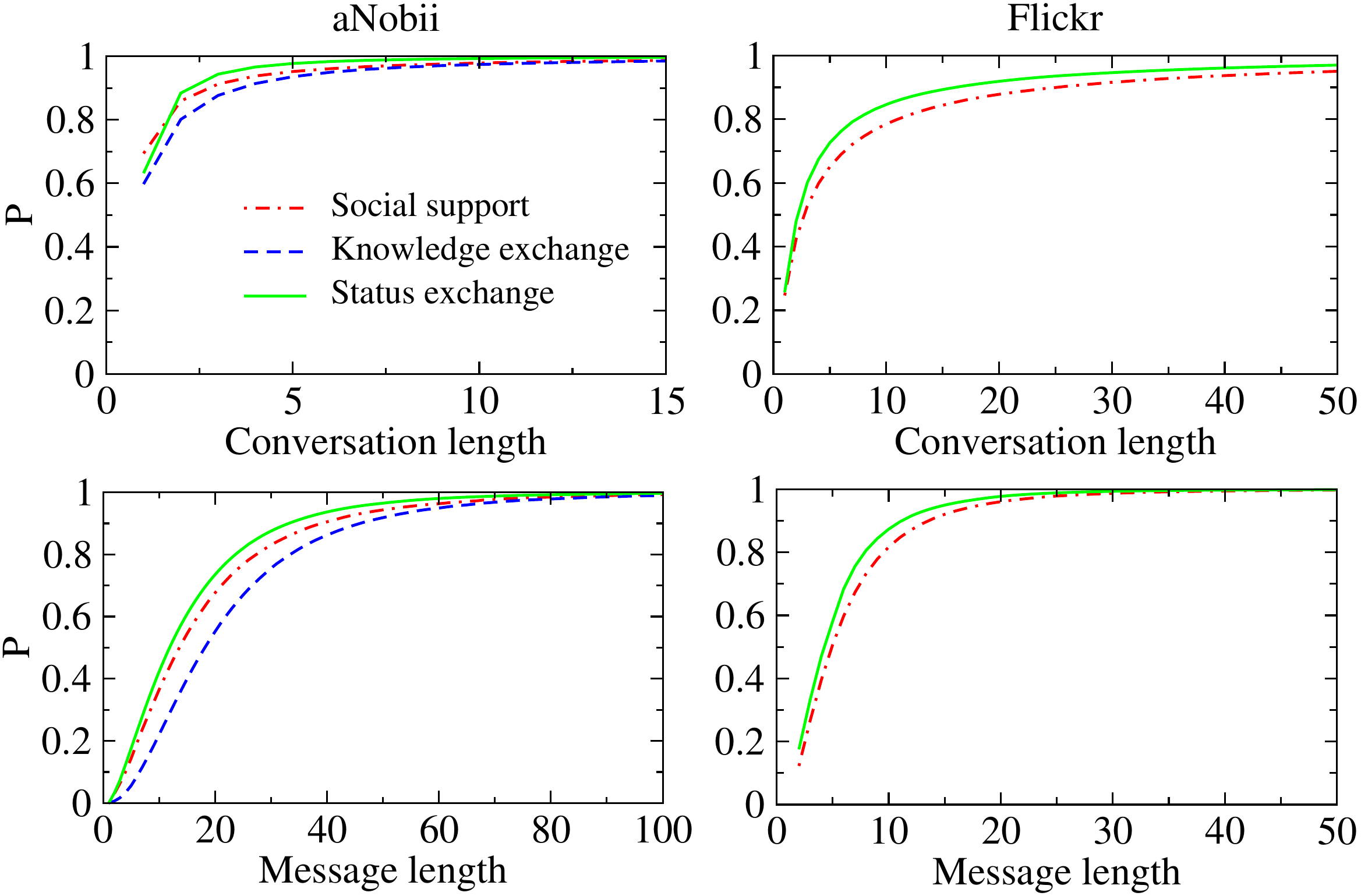}
 \footnotesize 
 \caption{Cumulative probability distributions of length of conversations (number of messages exchanges) and length of messages (number of tokens).}
 \label{fig:distr}
\end{figure}

\begin{table*}[t]
\begin{center}
    \footnotesize
    \begin{tabular}{c|c|c|ccc|cc|cc|c}
			                \multicolumn{1}{c}{}   &   & \textbf{Tie Share} & \multicolumn{3}{c|}{\textbf{Structural sim}} & \multicolumn{2}{c|}{\textbf{Intensity}} & \multicolumn{2}{c|}{\textbf{Sentiment}} & \textbf{Kinship}\\
			\cline{4-10}
		      \multicolumn{1}{c}{}  & \textbf{DoI}    & & \textbf{$\sigma_n$} & \textbf{$\sigma_g$} & \textbf{$\sigma_i$} & \textbf{$\langle conv_{len}\rangle$} & \textbf{$\langle msg_{len} \rangle$} &Intim. & Emo. &  \\ 
		  \hline
		  \multirow{3}{*}{aNobii} & \textit{Status}       & 0.48 & 0.045 & 0.062 & 0.041 & 2.13 & 16.32 & 0.026 & 0.033 & n/a \\
		  & \textit{Support}                              & 0.33 & 0.064 & 0.077 & 0.054 & 3.03 & 18.81 & 0.040 & 0.040 & n/a \\
		  & \textit{Knowledge}                            & 0.19 & 0.068 & 0.075 & 0.059 & 2.48 & 23.27 & 0.038 & 0.036 & n/a \\
			\hline
			\multirow{2}{*}{Flickr} & \textit{Status}       & 0.51 & 0.028 & 0.024 & 0.0011 & 8.83 & 6.26 & 0.370 & 0.393 & 0.049 \\
		  & \textit{Support}                              & 0.49 & 0.040 & 0.024 & 0.0013 & 12.70 & 7.35 & 0.410 & 0.440 & 0.057 \\
    \end{tabular} 
 \caption{Strength of ties connecting pairs of users, in terms of: i) Jaccard similarity $\sigma$ between their neighbors ($n$), the groups they are subscribed ($g$) and their items ($i$), books for aNobii and favorited photos for Flickr; ii) lenght of the conversation in terms of number of messages exchanged; iii) ratio of words belonging to the intimacy and emotions categories in the LIWC categories; iv) ratio of dyads reciprocally declaring a ``family'' or ``friend'' relation (available for Flickr only). The portions of messages per dyad belonging to each DoI, averaged over all the dyads, is also reported as \textit{tie share}.}
 \label{tab:tiestrength}
  \end{center}
\end{table*}

\subsection{Tie composition and strength}\label{sec:analysis:strength}

Our approach allows us to decompose a social link in the DoIs that constitute the communication between its endpoints. We study the proportion of different resources exchanged over a communication tie. The portions of messages per dyad belonging to each DoI, averaged over all the dyads, is reported in the ``tie share'' column in Table~\ref{tab:tiestrength}. In aNobii, status giving is the most frequent interaction. This finding is rather intuitive: status giving is predominant in very short messages -- the archetypal message in this context being ``nice library!'' -- and many messages are relatively short. In Flickr the proportion is very balanced instead, with no domain being predominant on average. This finding is consistent with recent studies on group interactions in Flickr that identify a dichotomy between \textit{social} and \textit{topical} interactions in the network~\cite{grabowicz13distinguishing}.

In Figure~\ref{fig:distr} we show the distributions of message length (in number of tokens) and conversation length for the three domains. This difference is not only helpful in characterizing the interactions but it also means that the DoI can potentially serve as an additional feature in the task of conversation length prediction~\cite{backstrom2013}.

Another important aspect that connects our work with previous studies in social link characterization is the measurement of the average strength of the link in the different DoI graphs. To understand whether differences in the strength of the ties hold between different domains, we adopt the framework presented by Gilbert and Karahalios~\cite{gilbert09predicting} based on Granovetter's definition of tie strength~\cite{Granovetter:1973}. Unlike in Gilbert's experiment, we do not have a crowdsourced ground truth about tie strength in our datasets. However, our aim here is not to determine which metrics are better proxies for tie strength, but to compare individual strength indicators across DoIs, to spot differences between them. We measure the strength based on three main families of metrics~\cite{gilbert09predicting}: \textit{structural similarity} (extent to which the tied individuals share common acquaintances and features), \textit{intensity} (duration of their interaction), and \textit{sentiment} (amount of words expressing intimacy and emotions). To quantify the sentiment dimension, we process the text of conversations using the English and Italian versions of the ``Linguistic Inquiry Word Count'' (LIWC) dictionary~\cite{pennbacker13secret}. LIWC is a dictionary that maps words into 72 categories, such as positive and negative emotional words, words implying cognitive processes, psychological constructs, and so on. To capture the notion of intimacy and emotion, we use the LIWC intimacy categories previously identified in~\cite{gilbert09predicting}. In addition, Flickr data allows us to investigate also the \textit{kinship} dimension, namely whether a social partners declared to be friends or member of the same family.

Results are presented in Table~\ref{tab:tiestrength}. In aNobii, ties in the Status Exchange network exhibit the lowest strength in terms of all indicators. Support and Knowledge networks have more similar features, with the Knowledge Exchange ties having slightly higher similarity in terms of books shared between the endpoints and the Social Support ties having longer conversations, on average. Flickr has consistent results for all the feature categories even if the signal is weaker for some structural indicators, as the structural similarity is substantially different only when considering the similarity of common friends. Also, the probability of people exchanging support being real life friends or kins is $16\%$ higher than for status. 

In summary, in both networks weaker ties tend to convey status giving and stronger ties either social support or knowledge.

\begin{figure}[t]
\center
 \includegraphics[width=\columnwidth]{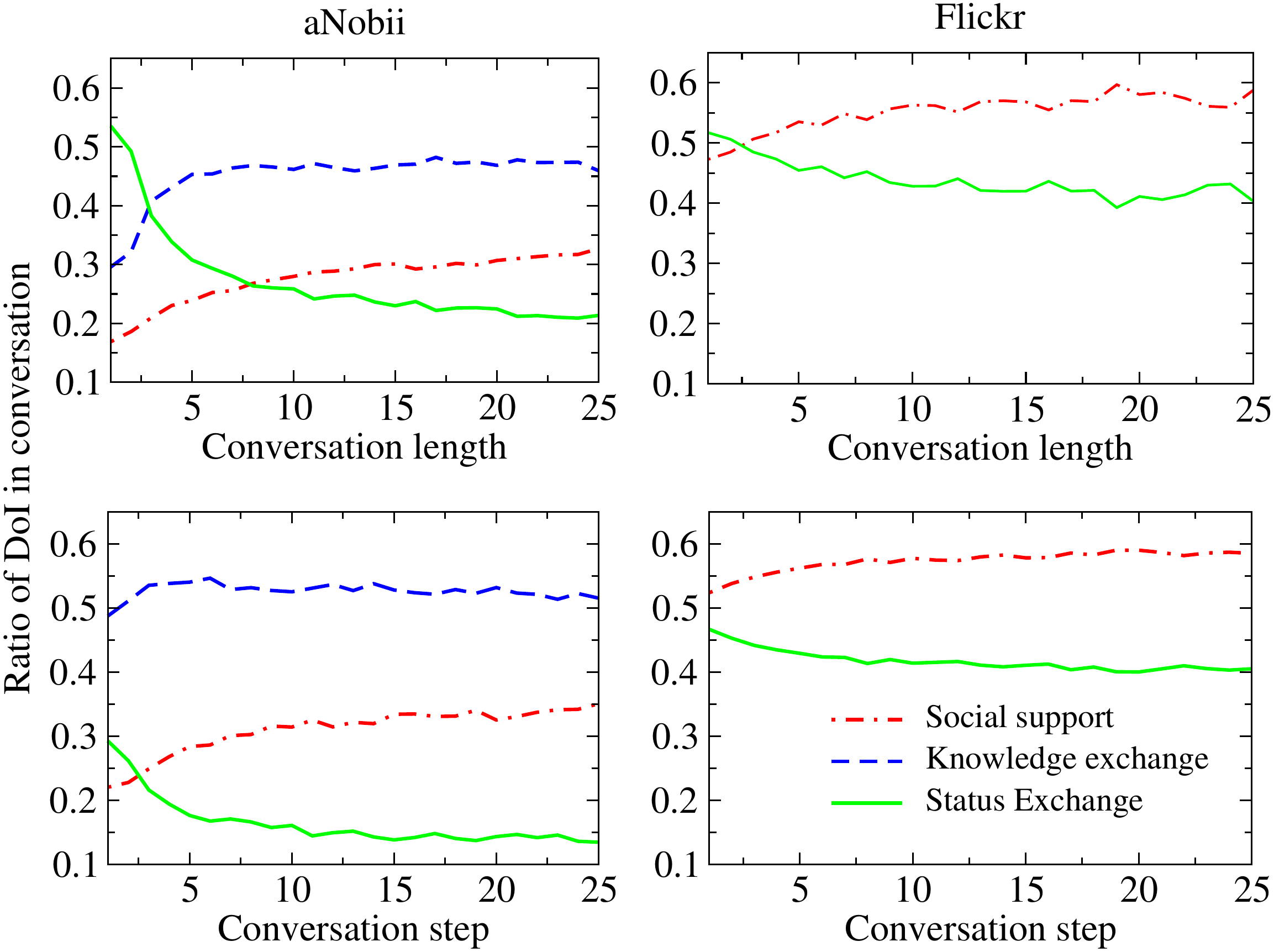}
 \footnotesize 
 \caption{Average proportion of messages belonging to each DoI for pairs of users with fixed conversation length (top) and after $n$ conversation steps (bottom).}
 \label{fig:vslength}
\end{figure}

\subsection{Tie evolution}
Intuitively, the role and importance of each domain in a dyadic relation could potentially change as the relationship evolves in time. To study the evolution of social processes along a tie after its creation, we compute across all the users the average ratio of messages belonging to each DoI in \textit{i)} conversations with different lengths and \textit{ii)} in messages belonging to the $n^{th}$ conversation step. Figure~\ref{fig:vslength} shows the dynamic of this evolution. Status exchange is particularly present in short conversations or, more in general, in the first stages of a conversation, after which the average tie moves to a mix of knowledge exchange and social support. It thus appears that status exchange serves to set the foundation for the future relationship, fading to the interactional background after the tie-formation stage. Interestingly, the same pattern (even if smoother) is found for Flickr, where status giving is predominant at the beginning and then slowly loses its importance. Even more surprisingly, in both datasets the status giving curve starts losing its predominance exactly after $3$ messages exchanged.

Even though highly-reciprocal status decreases as a conversation grows in length, reciprocity nonetheless tends to \textit{increase} in a relationship over time (Figure~\ref{fig:reciprocity}). This is a likely example of survival bias among social ties. Power-imbalanced relations, where only one individual provides resources and the other cannot reciprocate, are assumed to be more vulnerable to dissolution through the dependent actor's withdrawal from the relationship~\cite{Emerson:1962}. Thus, we are more likely to observe long conversations stemming from reciprocal relationships than from non-reciprocal ones.

\begin{figure}[t]
\center
 \includegraphics[width=0.7\columnwidth]{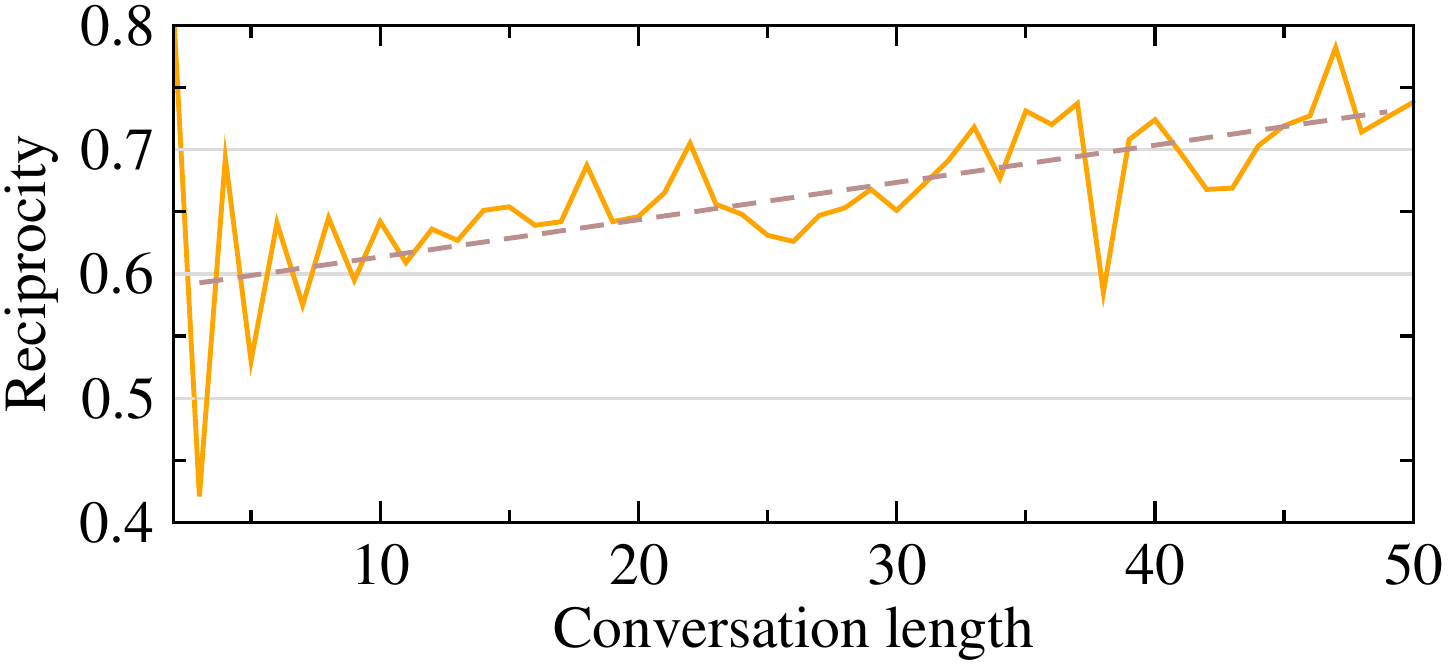}
 \footnotesize 
 \caption{Average reciprocity of interactions at fixed conversation length. Linear fitting is reported.}
 \label{fig:reciprocity}
\end{figure}

\subsection{Inequality and assortativity}

In our conception, inspired by the Social Exchange Theory, knowledge, status, and support can be considered as \textit{goods} generated by the social actors and exchanged between them. We investigate the way in which the exchange of such goods is distributed in the network. A common way to measure the \textit{social inequality}, i.e., the tendency of small circles of people to accumulate the vast majority of the global wealth, is to draw the Lorenz curve of any wealth indicator. The curve plots the proportion of the global wealth retained by the poorest $x\%$ of the population: the farther the curve is from the diagonal, the greater the inequality between individuals. In Figure~\ref{fig:lorenz} we plot the Lorenz curve by using the indegree (similar results are obtained with the in-strength) of each DoI graph as a proxy of wealth (e.g., number of alters giving status to ego) and we compute the Gini coefficient $G\in[0,1]$ as a quantitative measure of the inequality~\cite{gastwirth72estimation}. In general, the distribution of resources is very unequal in all the domains but in particular for the status giving, which has the highest Gini coefficient: in aNobii $G_{sta}=0.72$, $G_{sup}=0.69$, and $G_{kno}=0.68$ and in Flickr $G_{sta}=0.53$, $G_{sup}=0.43$). This supports the intuition that the status, more than other goods, tends to flow unidirectionally from low- to high-status individuals.

We investigate the social stratification also by measuring the in-in assortativity of the graphs, namely the tendency of individuals to connect with people with similar indegree~\cite{newman02assortative}. In Figure~\ref{fig:assortativity} we report the assortativity values for the three subgraphs and the full communication graph. To check the statistical significance of results, i) we compute the same values on randomly rewired versions of the graphs and ii) we compute the error on the assortativity estimation through jackknife resampling~\cite{newman02assortative}. Surprisingly, in aNobii all the assortativity absolute values tend to zero, meaning that the connectivity patterns in all the networks are very mixed. Status is the only DoI that tends to disassortativity, thus confirming the tendency of unidirectional status flow (i.e., people with higher status receiving status from people with lower status). The full communication network is disassortative as well because dominated by the signal of the Status DoI, which covers the highest number of edges (see Table~\ref{tab:doinets}). In Flickr, assortative patterns are more evident but, consistently with aNobii, status assortativity is lower than for social support, with a statistically significant difference.

\begin{figure}[t]
\center
 \includegraphics[width=\columnwidth]{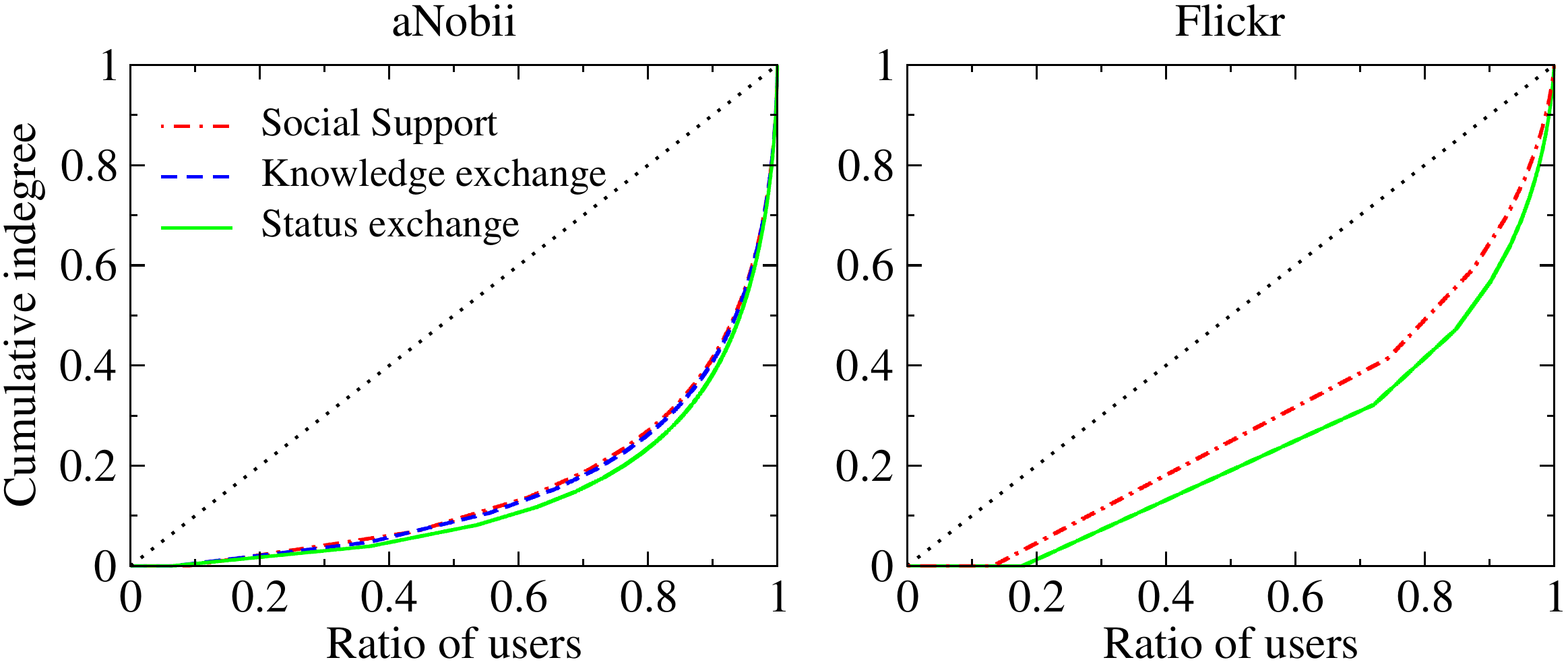}
 \footnotesize 
 \caption{Lorenz curve for the indegree (unweighted on the left and weighted on the right) in the three DoI graphs.}
 \label{fig:lorenz}
\end{figure}

\section{Discussion and Conclusions}

The methodology we propose has two immediate outcomes. First it provides an unsupervised way to discover the type of social exchange (e.g., status giving) that happens with dyadic passing of messages, in contrast with other methods that are able to capture the message's topic or sentiment. The accuracy of our approach in assigning messages to different domains is high, as assessed by human evaluators and consistently good in networks with direct user-to-user messaging (aNobii) as well as with item-mediated communication (Flickr). Last, it allows us to study the structure of the different interaction networks and to check our quantitative findings against well-established sociological theories. Among other findings, we verify that strong links in the communication network tend to convey either social support or knowledge, while weaker links convey more status giving. We also gain insights into the way ties evolve over time with status exchange gradually giving way to exchanges of knowledge or social support. Interestingly, the predominance of status exchange fades after $3$ message exchanges on average in both the datasets we tested.

\begin{figure}[t]
 \includegraphics[width=\columnwidth]{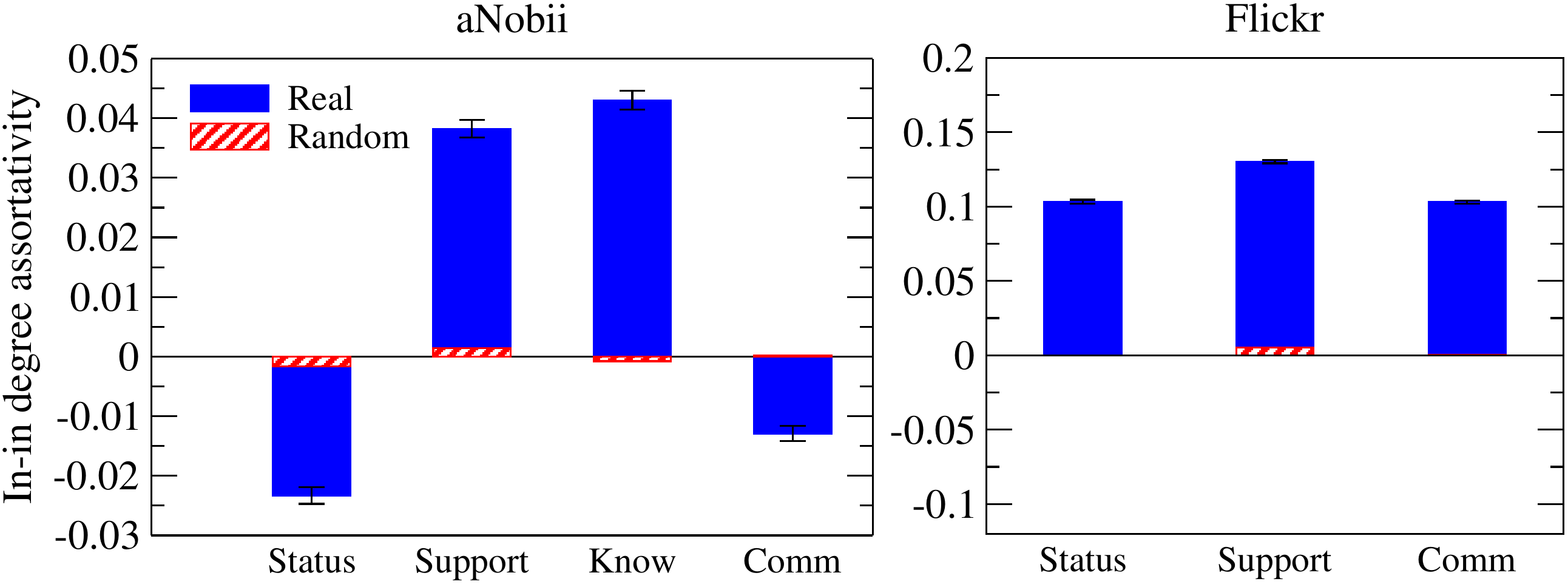}
 \footnotesize 
 \caption{In-in assortativity in the DoI graphs and in the full graphs. Errors estimated with jackknife resampling are always $<10^{-3}$ and are showed as error bars. Values computed on randomly rewired networks are shown with striped bars.}
 \label{fig:assortativity}
\end{figure}

The characterization of messages in terms of their type of social exchange opens a plethora of unexplored opportunities for applications, not limited to analytics. First is \textit{user profiling}: users engaged in conversations that are predominantly characterized by different DoIs would be presumably interested in different types of activities (e.g., socialization vs. item consumption). Second is \textit{link profiling}: dyads exchanging different social resources might react differently to signals. For example, when considering a process of information diffusion (e.g., diffusion of product ads via viral marketing), considering the knowledge, status or social support networks may yield very different results. Last, we see opportunities for the \textit{summarization} of social relationships. For example, Facebook's friendship page\footnote{\tiny\url{newsroom.fb.com/News/531/A-New-Look-for-Friendship-Pages}} displays a relationship between two connected users with a timeline of their shared experiences. Our tie decomposition in domains would allow a different way of summarizing a social link, e.g., ``based on their conversations, Alice and Bob's relationship has been made $30\%$ by knowledge exchange, $20\%$ by status giving and $50\%$ by social support.''

Our method has also some limitations that we plan to address in the future and that we summarize as follows.

\textbf{Supervised vs. unsupervised.} Our approach is fully unsupervised. This choice is motivated by the purpose of \textit{discovery} of the framework: detecting the domains of interaction in \textit{any} communication network. Supervised alternatives are possible. If a ground truth is available, a training set could be built from any set of features (textual, social, and so on). However, such approach would need i) an initial labeling effort, ii) to build different ground truth corpora for different domains, and iii) to know in advance the number and type of resources that are exchanged in the network. Our approach is free from these constraints and therefore more general. We plan to explore combinations of supervised and unsupervised approaches for a classification of messages on the fly.

\textbf{Clustering alternatives.} We used NMF in the message bucketing stage ($\S$\ref{sec:bucketing}) and Spinglass as community detection algorithm in the phase of DoI extraction ($\S$\ref{sec:community}), but a plethora of alternatives for clustering and community detection are available. We also conducted experiments using Latent Dirichlet Allocation (LDA) and Fuzzy K-Means in alternative to NMF and we found analogous results. We plan as a future work to experiment more community detection algorithms in alternative to Spinglass.

\textbf{Message bucketing.} The bucketing phase groups messages by the similarity of their bags of words but other types of aggregation to better capture the semantics of messages would be possible. We partially address this point by giving in input to the clustering also bi-grams and tri-grams, that are needed to account for associations of words with slightly more complex meaning. Also clustering messages by their sentiment would be an interesting extension.

\textit{Concluding remarks.} The representation of a social tie as a sequence of individual exchanges naturally leads one to the idea of understanding social ties as strings of interactions. With this understanding, we can use insights from theoretical Computer Science to establish the computational properties of social rituals. Indeed, this idea has already been leveraged by DeDeo~\citeyear{DeDeo:2012}, who gives evidence of the insufficiency of finite-state machines for the description of social interactions. The ultimate goal of such analysis is the unpacking of ``culture'' as a formal, computational concept. If we see social ties as interactional sequences, then we may understand the Domains of Interaction we discover as the ``grammar of society''~\cite{bicchieri06grammar} -- in other words, the bits of ``source-code'' that prescribe how individuals are to act in a certain situation. We hope our work provides yet another step towards a truly computational understanding of human societies.

\section*{Acknowledgments}
\small
This research is supported by EU Community's Seventh Framework Programme FP7/2007-2013 under the ARCOMEM and Social Sensor projects, by the Spanish Centre for the Development of Industrial Technology under the CENIT program, project CEN-20101037 ``Social Media'', and by Grant TIN2009-14560-C03-01 of the Ministry of Science and Innovation of Spain. R. Schifanella was partially supported by the Yahoo FREP grant.

\balance
\small
\bibliographystyle{abbrv}

\end{document}